# Low-complexity prediction of complex-valued sequences using a novel "residual-as-prediction" method

Thomas Tetzlaff
*Intel Corporation*
thomas.a.tetzlaff@intel.com

**Abstract**

*A method of prediction is presented to aid compression of sequences of complex-valued samples. The focus is on using prediction methods to reduce the average magnitude of residual values after prediction (not on the subsequent compression of the residual sequence). The prediction method has low computational complexity, so as to keep power consumption in implementations of the method low. The new method presented here applies specifically to sequences that occupy a significant percentage of the sampling bandwidth; something that existing, simple prediction methods fail to adequately address. The new method, labeled "residual-as-prediction" here, produces residual sequences with reduced mean magnitude compared to the original sequence, even for sequences whose bandwidth is up to 85% of the sampling bandwidth.*

## 1. Introduction

The organization of the paper is as follows: First, a simple time correlation is described which is a typical approach used particularly for sequences that occupy a low bandwidth compared to the sampling frequency. Then, a method of prediction is introduced that provides a reduction in residual amplitude even when the occupied bandwidth is as high as 85%. Next, enhancements are detailed to provide improved amplitude reduction in the residual sequence for occupied bandwidths between 20 and 70 percent. Finally, a summary of the results are provided, including some notes on implementation.

## 2. Time-based Correlation

The topic of obtaining predictions for a time series has been well covered [1-3]. One very simple form of sequence prediction is to use an expected value of the unit-delay auto-correlation, which is labeled as "time correlation" here. Consider a sequence of samples, $samp$, at time index $t$. Then the expected value of the time correlation of the complex sequence is given by:

$$timecorr = \frac{\sum (samp_t \cdot \overline{samp_{t-1}})}{\sum (samp_t \cdot \overline{samp_t})} \qquad (1)$$

In practice, one can choose to update the time correlation periodically as well as calculate the value over a specified number of samples. Given the time correlation and a current value in the sequence, a prediction for the next value in the sequence and the residual sequence are given by:

$$prediction_{t+1} = timecorr \cdot samp_t \qquad (2)$$

$$residual_{t+1} = samp_{t+1} - prediction_{t+1} \qquad (3)$$

From the residual sequence and time correlation, the original sequence can be reconstructed. For a single-frequency tone, the time correlation will have a magnitude of 1 and a phase equal to the rotation per sample of the tone. The prediction in this case is perfect after the initial sample, and the residual sequence will be a sequence of zeros for t>0. As the bandwidth of the sequence to be compressed increases, the magnitude of the time correlation will decrease, approaching 0 as the bandwidth of the sequence to compress approaches the sampling bandwidth. The phase of the complex-valued time correlation will be equal to an energy-average rotation-per-sample for the original sequence.

## 3. Residual-as-prediction Method

To address sequences that are a substantial fraction of the sampling frequency, this paper introduces a new, simple method for prediction: using the previous residual value as a prediction for the next value. This simple method is labeled as "residual-as-prediction" here. In terms of formulas, the residual values are always calculated as shown in eq. 3. However, now the prediction are calculated as:

$$pred_{t+1} = (1-\epsilon) \cdot residual_t \qquad (4)$$

To dampen oscillations, a small factor epsilon is introduced. In practice epsilon is usually picked to be in the range 0.005 and 0.03. Also note that the prediction should be quantized to match the quantization of the original sequence. In addition, scaling can be introduced to provide a mechanism for lossy compression. If no scaling is applied, the residual sequence can be used to reconstruct the original sequence without loss. The residual value at time t=0 is set to the value of the original sequence *samp* at time t=0. Finally, equation [4] assumes that the signal energy averages to 0. If that is not the case, the prediction is rotated by the phase of the time correlation, resulting in equation [5].

$$pred_{t+1} = (1-\epsilon) \cdot \frac{timecorr}{|timecorr|} \cdot residual_t \qquad (5)$$

Note that the term involving "timecorr" can be obtained by accumulating phase changes for samples by quadrants or octants to avoid any multiplication. For a single tone, the magnitude of the residual values settles to a value of 0.5 times the magnitude of the original tone. In this case, it is clear that the common time-correlation based approach is better. However, as the bandwidth of the target waveform increases, the mean magnitude of the residual values grows much slower than the above approach. It provides a general reduction in magnitude all the way out to a target waveform bandwidth that is 85% of the sampling bandwidth. Figure (1) illustrates the difference in residual values for the two approaches as the bandwidth of the original sequence increases. For signal bandwidths more than 30% of the sampling bandwidth, the residual-as-prediction approach is better than the time-correlation approach. Note that the mean magnitude of the original sequence is one.

Once the magnitude of the residual signal is more than 90% of the original signal, there is almost no benefit in compressing the residual signal compared to the original. As such, while the time correlation method of prediction is good for small signal bandwidths, it is ineffective once the signal bandwidth is more than two-thirds of the sample bandwidth. Note that a residual sequence whose amplitude is 90% of an original sequence with 10-bit real and imaginary parts would still only provide approximately 98.3% compression factor. Compression factor is defined as compressed size divided by original size.

To give an explicit example of the residual-as-prediction method, consider a signal sequence that goes 10,10,10,10… forever. So, this is a DC sequence. If the starting prediction is 20 and epsilon is zero, then the residual value sequence will go: -10, 20, -10, 20… Obviously, the residual sequence has higher mean energy and magnitude than the original sequence – it utterly fails at helping to compress the sequence. This is why a small factor is used to relax the predictions and damp out the oscillating behavior. With a small non-zero value for epsilon, the prediction will relax to 5 (or, within the small factor of it), and the residual sequence becomes: 5,5,5,5… Thus, we get a sequence with half the mean magnitude of the original sequence.

## 4. Frequency Response

As a means of explaining the residual sequence behavior, the frequency response of the residual sequences is examined. In figure (2), the magnitude of the frequency components is plotted for an original sequence and 3 different residual sequences using time correlation. The original sequence uses 80% of the sampling bandwidth, is centered on DC, and all frequency components have unit magnitude. There is no DC component, however. The light blue line shows the response of time correlation-based residual for an original sequence using 20% of the sampling bandwidth. In this case, the magnitude of the time correlation will be near 1. The components nearest DC are close to zero, but the response grows rapidly as one moves away from DC. The green line shows the response when the original sequence uses 50% of the sampling bandwidth. In this case, the magnitude of the time correlation is reduced, such that the response of near-DC components is no longer close to zero. But, this minimizes the impact of the large frequency components of the residual, which are now larger than the original sequence. Finally, the magenta line shows the response when the original sequence uses 80% of the sampling bandwidth. Now, the time correlation magnitude is close to zero. So, the components near DC, which would've been very small, are now moving towards one. And, the high frequency components, which would've been very large, are also moving towards one. The overall effect is that the mean magnitude of the residual sequence is close to the original sequence. This figure shows why, in figure 1, the residual values grow rapidly when using time-correlation as the signal bandwidth increases.

In figure (3), the magnitude of the frequency components is plotted for an original sequence using 50% of the sampling bandwidth as well as 3 residual sequences using residual-as-prediction. The green line shows the response for residual-as-prediction as described in the previous section. Here, the components near DC have half the magnitude of the original sequence. However, the growth in magnitude as one moves away from DC is much slower than for time correlation. There is also the presence of a +/- sampling-frequency/2 component. This is the oscillating component controlled by the epsilon factor. Smaller epsilons increase the magnitude of this component. The light blue and magenta lines are labeled 2-pass and 3-pass, which will be describe more in the next section.

## 5. Enhancements to residual-as-prediction

Looking at the frequency response of the residual sequence when using residual-as-prediction, it simply has the same frequency components as the original sequence, except for a small amount of energy near positive and negative of half the sampling rate. This energy can be adjusted by changing the epsilon value – larger epsilon values will reduce this energy present in the residual sequence, but will increase the response at other frequencies. Due to this frequency distribution, it is clear that the residual-as-prediction can be repeated using the residual sequence as a new "original" sequence, creating a 2-pass residual sequence. And this can be repeated again and again. However, due to the amplification of the Nyquist-rate energy as well as the increased computational complexity, the 3-pass option is used here as a somewhat practical limit to the number of passes. Figure (4) adds the 2-pass and 3-pass results to the original figure (1).

In practice, there are a few implementation options. For example, if most of the sequences encountered are ones occupying 74% to 85% of the sampling bandwidth, one might choose to forego high compression for rarely observed low bandwidth sequences, instead choosing to go with 1-pass residual-as-prediction always.

Alternatively, if varying bandwidths are expected, one could calculate all residual sequence options, picking the best one at any given time: time-correlation for sequences occupying less than 8% of the sampling bandwidth, 3-pass residual-as-prediction when bandwidth occupation is between 8% and 74%, 1-pass when bandwidth usage is 74% to 85%, and bypassing prediction otherwise.

Note that while the compression of the residual sequence could be lossy or lossless, these results are looking at the mean magnitudes of the lossless residual sequence. In addition, the simulations used floating point values, but this doesn't impact the applicability to fixed point values. One simply has to quantize prediction values.

## 6. Summary

A method of prediction is presented in which the residual is defined as the difference between the original sample value and the predicted value. In this work, we have shown that using this residual value as a prediction of the next sample (residual-as-prediction) offers significant benefits for complex-valued sequence prediction compared to time correlation. This applies to a wide range of signals, occupying between 8% to 85% of the sample bandwidth. While results have focused on reduction in magnitudes of the residual sequence, this reduction in magnitude has led to improved compression performance.

All simulation code which was used to produce the figures is listed in the appendix.

## 7. Figures

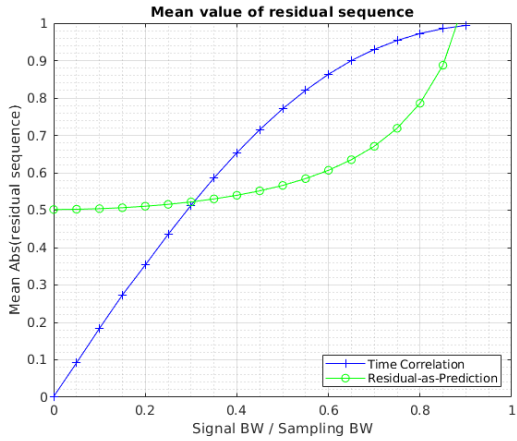

*Figure 1: Magnitude of residual sequences for different prediction methods*

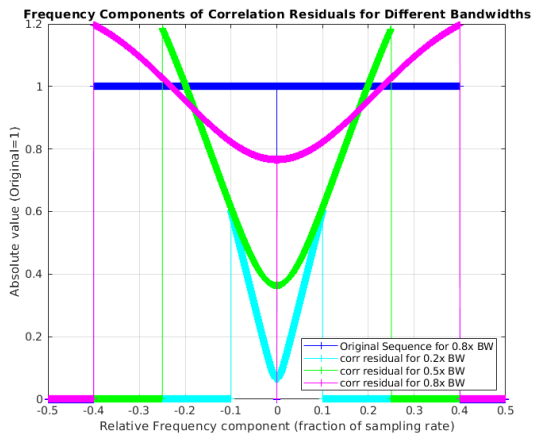

*Figure 2: Residual Sequence Frequency Response (Time Correlation)*

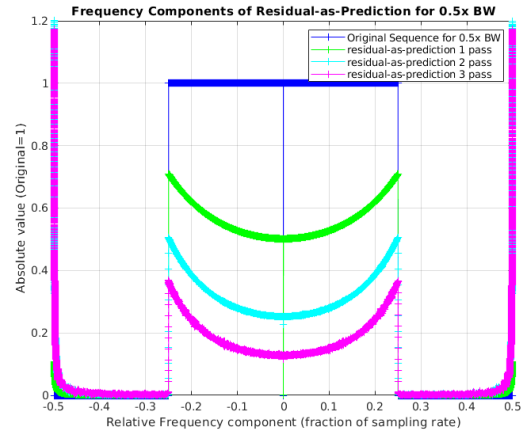

*Figure 3: Residual Sequence Frequency Response (residual-as-prediction)*

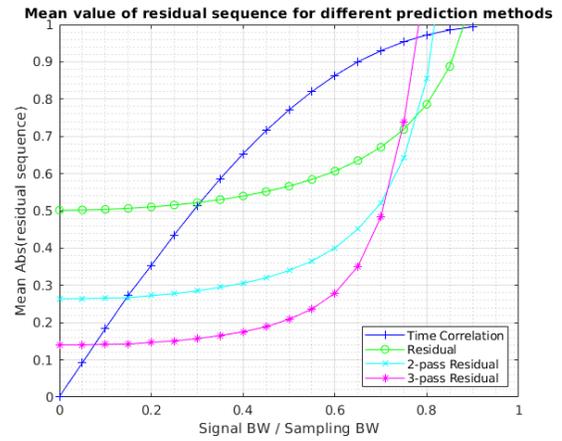

*Figure 4: Residual Magnitudes vs. Signal Bandwidth*

## 8. Appendix

Program listings used to generate figures in this paper.

# 8.1. Matlab listing of all functions

```matlab
%Copyright (c) 2019, Intel Corporation
%
%This program is free software; you can redistribute it and/or modify it
%under the terms and conditions of the GNU General Public License,
%version 2, as published by the Free Software Foundation.
%
%This program is distributed in the hope it will be useful, but WITHOUT
%ANY WARRANTY; without even the implied warranty of MERCHANTABILITY or
%FITNESS FOR A PARTICULAR PURPOSE.  See the GNU General Public License for
%more details.

% create plots of residual averages for correlation and residual-as-prediction
% correlation - epsilon=0.01 for small ratios, use full correlation otherwise
ep_timecorr = 0.01;
ep_arg = ep_timecorr;

% epsilon and saturation for 1-pass residual-as-prediction
ep1 = [0.007];
sat1 = [2.75];

% epsilon and saturation for 2-pass
ep2 = [0.012 0.01];
sat2 = [2.5 2.2];

% epsilon and saturation for 3-pass
ep3 = [0.015 0.03 0.01];
sat3 = [2.4 1.4 0.8];

num_trials = 10;

idx = 1;
residual_table=[];
rat_pts = (0:0.05:0.90);
for ratio=rat_pts
  residual_totals = zeros(1,4);
  for trials=1:num_trials
    if ( ratio >= 0.2 )
      ep_arg = 0; %use full sequence correlation
    end
    samp_td = pred_genseq( ratio/2, 65536, 1 );
    residual_corr = correlation_pred(samp_td, ep_arg);
    residual_1pass = residual_as_prediction(samp_td, ep1, sat1);
    residual_2pass = residual_as_prediction(samp_td, ep2, sat2);
    residual_3pass = residual_as_prediction(samp_td, ep3, sat3);

      residual_totals = residual_totals + mean(abs([residual_corr(:) residual_1pass(:) residual_2pass(:) residual_3pass(:)]));

  end
  result_table(idx,:) = residual_totals / num_trials;
  idx = idx+1;
end

figure(2);
plot(rat_pts,result_table(:,1),'-+b',rat_pts,result_table(:,2),'-og',rat_pts,result_table(:,3),'-xc',rat_pts,result_table(:,4),'-*m');
axis([0 1 0 1]);
grid on;
grid minor;
legend('Time Correlation','Residual','2-pass Residual','3-pass Residual','location','southeast');
xlabel('Signal BW / Sampling BW');
ylabel('Mean Abs(residual sequence)');
title('Mean value of residual sequence for different prediction methods');

figure(1);
plot(rat_pts,result_table(:,1),'-+b',rat_pts,result_table(:,2),'-og');
axis([0 1 0 1]);
grid on;
grid minor;
legend('Time Correlation','Residual-as-Prediction','location','southeast');
xlabel('Signal BW / Sampling BW');
ylabel('Mean Abs(residual sequence)');
title('Mean value of residual sequence');

%now, create plots for frequency responses
len = 65536*4;
[samp_td2 normalization2] = pred_genseq( 0.2/2, len, 0 );
residual_corr2 = correlation_pred( samp_td2, 0 );
[samp_td5 normalization5] = pred_genseq( 0.5/2, len, 0 );
residual_corr5 = correlation_pred( samp_td5, 0 );
[samp_td8 normalization8] = pred_genseq( 0.8/2, len, 0 );
residual_corr8 = correlation_pred( samp_td8, 0 );
residual_1pass = residual_as_prediction(samp_td5, ep1, sat1);
residual_2pass = residual_as_prediction(samp_td5, ep2, sat2);
residual_3pass = residual_as_prediction(samp_td5, ep3, sat3);

figure(3);
xaxis = (0:length(samp_td2)-1)/length(samp_td2) - 0.5;
orig = abs(fftshift(fft(samp_td8))) * normalization8;
orig5 = abs(fftshift(fft(samp_td5))) * normalization5;
tmp2 = abs(fftshift(fft(residual_corr2))) * normalization2;
tmp5 = abs(fftshift(fft(residual_corr5))) * normalization5;
tmp8 = abs(fftshift(fft(residual_corr8))) * normalization8;
res1 = abs(fftshift(fft(residual_1pass))) * normalization5;
res2 = abs(fftshift(fft(residual_2pass))) * normalization5;
res2mm = movmean(res2,10);
res3 = abs(fftshift(fft(residual_3pass))) * normalization5;
res3mm = movmean(res3,10);

plot(xaxis,orig,'-+b',xaxis,tmp2,'-+c',xaxis,tmp5,'-+g',xaxis,tmp8,'-+m');
legend('Original Sequence for 0.8x BW','corr residual for 0.2x BW','corr residual for 0.5x BW','corr residual for 0.8x BW','location','SouthEast');
xlabel('Relative Frequency component (fraction of sampling rate)');
ylabel('Absolute value (Original=1)');
title('Frequency Components of Correlation Residuals for Different Bandwidths');
grid on;

figure(4);
plot(xaxis,orig5,'-+b',xaxis,res1,'-+g',xaxis,res2mm,'-+c',xaxis,res3mm,'-+m');
axis([-0.5 0.5 0 1.2]);
legend('Original Sequence for 0.5x BW','residual-as-prediction 1 pass','residual-as-prediction 2 pass','residual-as-prediction 3 pass','location','NorthEast');
xlabel('Relative Frequency component (fraction of sampling rate)');
ylabel('Absolute value (Original=1)');
title('Frequency Components of Residual-as-Prediction for 0.5x BW');
grid on;

function [samp_td normalization] = pred_genseq(frac_pos_buckets,seq_len, norm_mag)

%function [samp_td] = pred_genseq(frac_pos_buckets,seq_len,norm_mag)
%
%generate unit-magnitude frequency components for the specified number
%of frequency slots both positive and negative side of DC
%if 1, only the single positive slot is picked

%Author: Thomas A. Tetzlaff

%Copyright (c) 2019, Intel Corporation
%
%This program is free software; you can redistribute it and/or modify it
%under the terms and conditions of the GNU General Public License,
%version 2, as published by the Free Software Foundation.
%
%This program is distributed in the hope it will be useful, but WITHOUT
%ANY WARRANTY; without even the implied warranty of MERCHANTABILITY or
%FITNESS FOR A PARTICULAR PURPOSE.  See the GNU General Public License for
%more details.

if ( frac_pos_buckets == 0 )
   num_pos_buckets = 1;
   num_neg_buckets = 0;
else
   num_pos_buckets = floor( frac_pos_buckets * seq_len );
   num_neg_buckets = num_pos_buckets;
end

if ( num_pos_buckets > (seq_len/2) )
   disp 'Error: too many positive frequency slots specified'
end

samp_fd = zeros(1,seq_len);
samp_fd(2:num_pos_buckets+1) = exp(1i*2*pi*rand(1,num_pos_buckets));
samp_fd(end-num_neg_buckets+1:end) = exp(1i*2*pi*rand(1,num_neg_buckets));

samp_td = ifft(samp_fd);
if ( norm_mag == 1 )
   normalization = mean(abs(samp_td)); %normalize magnitude
else
   normalization = sqrt(mean(samp_td .* conj(samp_td))); %normalize power
end
samp_td = samp_td / normalization;
end %end function

function [residual] = correlation_pred(samp_td, epsilon);
%function [residual] = correlation_pred(samp_td, epsilon);
%
%samp_td is a sequence of complex samples
%epsilon is small IRR factor for adaptive time correlation
% if epsilon is zero, the time correlation for the entire sequence
% is used as the prediction without any degradation
%
%residual is the output sequence of residual values using
%time-correlation
%

%Author: Thomas A. Tetzlaff
%
%Copyright (c) 2019, Intel Corporation
```

```matlab
len_seq = length(samp_td);
pred = 0;

full_corr = mean(samp_td(2:end) .* conj(samp_td(1:end-1))) / ...
            mean(samp_td(1:end-1) .* conj(samp_td(1:end-1)));
if ( epsilon == 0 )
   pred = full_corr;
else
   pred = 0;
end
residual = zeros(1,len_seq);
residual(1) = samp_td(1);
for i=2:len_seq
   residual(i) = samp_td(i) - pred*samp_td(i-1);
   % adding a small value to avoid 0 in denominator
   pred = (1-epsilon)*pred + epsilon*(samp_td(i) * conj(samp_td(i-1))) / ...
         (samp_td(i-1) * conj(samp_td(i-1))+1e-40);
end
end %end function

function [residual] = residual_as_prediction(samp_orig,epsilon,abs_threshold)
%
%function [residual] = residual_as_prediction(samp_orig,epsilon,abs_threshold)
%
%Inputs:
% samp_orig is the original sequence (ideally, energy is centered on DC)
%
% epsilon is a row vector of small (<<1) reduction factors for each pass
%    the number of elements in epsilon determines the number of passes
% abs_threshold is a row vector of thresholds for maximum prediction
%    values on each pass. This limits the range of residual sequences
%    It must have the same number of elements as epsilon
%
%Outputs:
% residual is the residual-as-prediction residual sequence for the Nth pass
%
```

```matlab
len_seq = length(samp_orig);
num_pass = length(epsilon);

residual_pass_mat = zeros(num_pass+1,len_seq);
residual_pass_mat(1,:) = samp_orig;
%samp_orig(1) is the starter value.  Can recover all from residual 1,2, or 3
residual_pass_mat(:,1) = samp_orig(1)*ones(num_pass+1,1);
prediction_pass = zeros(1,num_pass);

prediction_pass(1) = samp_orig(1);  %initial prediction for the N passes
for i=2:len_seq
   for pass=1:num_pass
      tmp = residual_pass_mat(pass,i) - prediction_pass(pass);
      residual_pass_mat(pass+1,i) = tmp;
      tmp = tmp * (1-epsilon(pass));
      if ( abs(tmp) > abs_threshold(pass) )
         tmp = tmp * abs_threshold(pass) / abs(tmp);
      end
      prediction_pass(pass) = tmp;
   end
end

residual = residual_pass_mat(end,:);
end %end function

function [samp_td] = decomp_residual(residual,epsilon,abs_threshold)
%
%function [samp_td] = decomp_residual(residual,epsilon,abs_threshold)
%
% residual is the list of residual values (can be pass 1, pass 2, or pass 3)
% epsilon is a vector with an entry for each pass of residual-as-prediction done
%   length(epsilon) will give the pass number used for the residual values
%   epsilon values must be the same used in compression
% abs_threshold is a vector of absolute value thresholds.  Again, these must
%   be the same as those used in compression
%   These are small IRR factors to limit oscillations
%
% samp_td is the returned decompressed I/Q values
%
```

```matlab
num_passes = length(epsilon);

pred_vals = zeros(num_passes);
pred_vals(1) = residual(1);
tmp_pred = zeros(num_passes+1);

samp_td = zeros(1,length(residual));
samp_td(1) = residual(1);

for i=2:length(residual)
   tmp_pred(num_passes+1) = residual(i);
   for pass=num_passes:-1:1
      tmp_pred(pass) = tmp_pred(pass+1) + pred_vals(pass);
      pred_vals(pass) = tmp_pred(pass+1) * (1-epsilon(pass));
      if ( abs(pred_vals(pass)) > abs_threshold(pass) )
         pred_vals(pass) = pred_vals(pass) * ...
            abs_threshold(pass)/abs(pred_vals(pass));
      end
   end
   samp_td(i) = tmp_pred(1);
end
end %end function
```